\documentclass[11pt]{article}
\usepackage{graphicx}
\usepackage{amssymb}
\usepackage{lineno}
\usepackage{lscape}

\newcommand{\scmokka}{\textsc{Mokka}}

\newcommand{\scmarlin}{\textsc{Marlin}}
\newcommand{\fivefive}{5\,$\times$\,5\,mm$^2$} 
\newcommand{\fortyfive}{45\,$\times$\,5\,mm$^2$} 
\newcommand{\tenten}{10\,$\times$\,10\,mm$^2$} 

\newcommand {\altten}{\ensuremath{\textrm{alt}_{10}}}
\newcommand {\altfive}{\ensuremath{\textrm{alt}_{5}}}
\newcommand {\mfivefive}{5$\times$5}
\newcommand {\mfortyfive}{45$\times$5}
\newcommand {\mfifteenfifteen}{15$\times$15}
\newcommand{\masspizero}{M$_{\pi^0}$}

\title{\bf Extraction Method of Fine Granular Performance from  Scintillator Strip Electromagnetic Calorimeter}
\author{Katsushige Kotera$^a$\footnote{Corresponding author (coterra@azusa.shinshu-u.ac.jp)}, Daniel Jeans$^b$, Akiya Miyamoto$^c$, and Tohru Takeshita$^a$
\vspace{.3cm}\\
$^a${\it Department of Physics, Shinshu University,}\\
{\it 3-1-1 Asahi, Matsumoto, Nagano 390-8621, Japan}\\
$^b${\it Department of Physics, Graduate School of Science, The University of Tokyo,}\\
{\it 7-3-1 Hongo, Bunkyo-ku, Tokyo 113-0033, Japan}\\
{$^c${\it High Energy Accelerator Research Organization (KEK),}}\\
{\it 1-1 Oho, Tukuba, Ibaraki 305-0801, Japan}}
\begin{document}


\maketitle 

\abstract

We describe an algorithm which has been developed to extract fine granularity information 
from an electromagnetic calorimeter with strip-based readout.
Such a calorimeter, based on scintillator strips, is being developed to apply particle flow reconstruction to future experiments in high energy physics.
Tests of this algorithm in full detector simulations, using strips of size  \fortyfive\, show that the performance is close to that
of a calorimeter with true \fivefive\ readout granularity.
The performance can be further improved by the use of \tenten\ tile-shaped layers interspersed between strip layers.

\section{Introduction}

In the experiments being designed for next generation electron-positron colliders, the particle flow approach (PFA)\cite{cite:MarkT,cite:JCB} 
is the leading candidate to achieve the 
excellent jet energy resolution required to fully exploit the possibilities afforded by the colliders' well-defined initial and clean final states.
In PFA, the energy of charged particles is measured by the tracking system, which has a much better momentum resolution than the energy resolution of calorimeters. 
The calorimeters are used to estimate the energy of only neutral particles.
In order to apply this approach, the calorimetric showers of each particle  must be individually reconstructed.
The granularity of calorimeter readout is therefore a key issue.

As an example, the sampling electromagnetic calorimeter (ECAL) being designed for the International Large Detector 
(ILD, a detector being designed for use at the International Linear Collider (ILC) \cite{cite:DBD}) is required to have 
a lateral segmentation of \fivefive\ and 20 - 30 longitudinal samplings, giving a total of $\sim10^8$ readout channels.
One technology being developed to achieve this high calorimeter granularity is based on  plastic scintillator strips individually read out by miniature photon detectors,
for example  pixelated photon detectors (PPD). 

The use of long scintillator strips rather than tiles of size \fivefive\ makes the design of such an ECAL more feasible, and reduces its cost, due to the 
reduced number of readout channels. Successive ECAL layers have orthogonally aligned strips, giving an effective granularity
close to the strip width.
The CALICE collaboration has developed and constructed ECAL prototypes based on this technology, using scintillator strips of length 45\,mm and width 5\ or 10\,mm,
individually read out by PPDs
\cite{cite:DESYTB, cite:Granada}.

This paper presents a reconstruction method which can be used to extract close to \fivefive\ effective granularity from such long scintillator strips, 
and reports on measurements of its performance using events fully simulated in ILD.
Details of this detector are given in the next section,
the reconstruction procedure is explained in section \ref{section:ssa}, and section \ref{section:calibration} describes the calibration procedure.
The position resolution achieved by this method is reported in section \ref{section:position}.
Particle separation abilities are reported on and discussed in section \ref{section:separation}, and  
the achieved jet energy resolution in two-jet events is reported and discussed in section \ref{section:JER}.
Finally we discuss the results in section\,\ref{section:discussion} and summarize this study in section\,\ref{section:summary}.

\section{Detector model} 

\begin{figure}[h!] 
\begin{center}
\includegraphics[width=.45\textwidth]{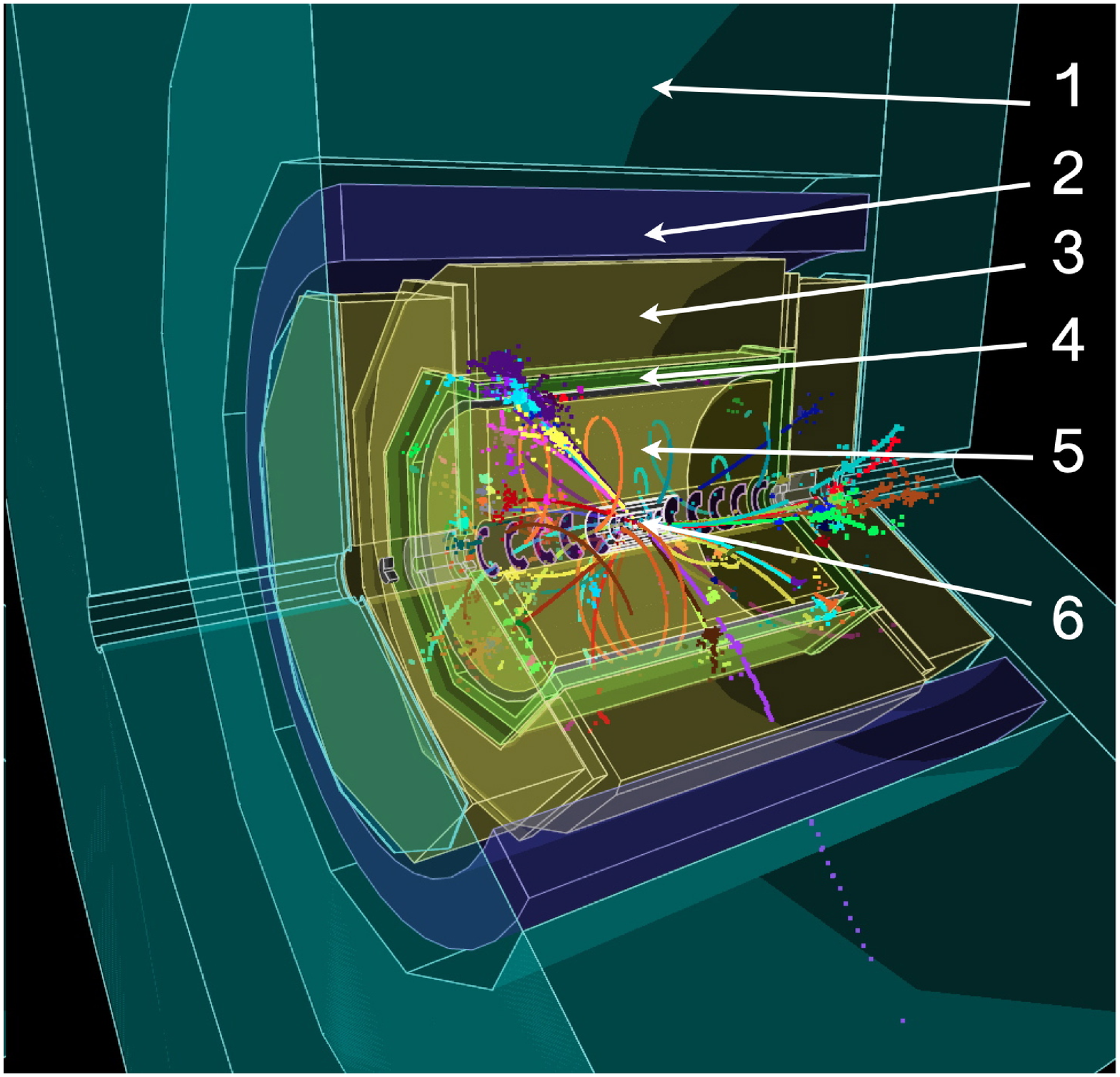}
\includegraphics[width=.4\textwidth]{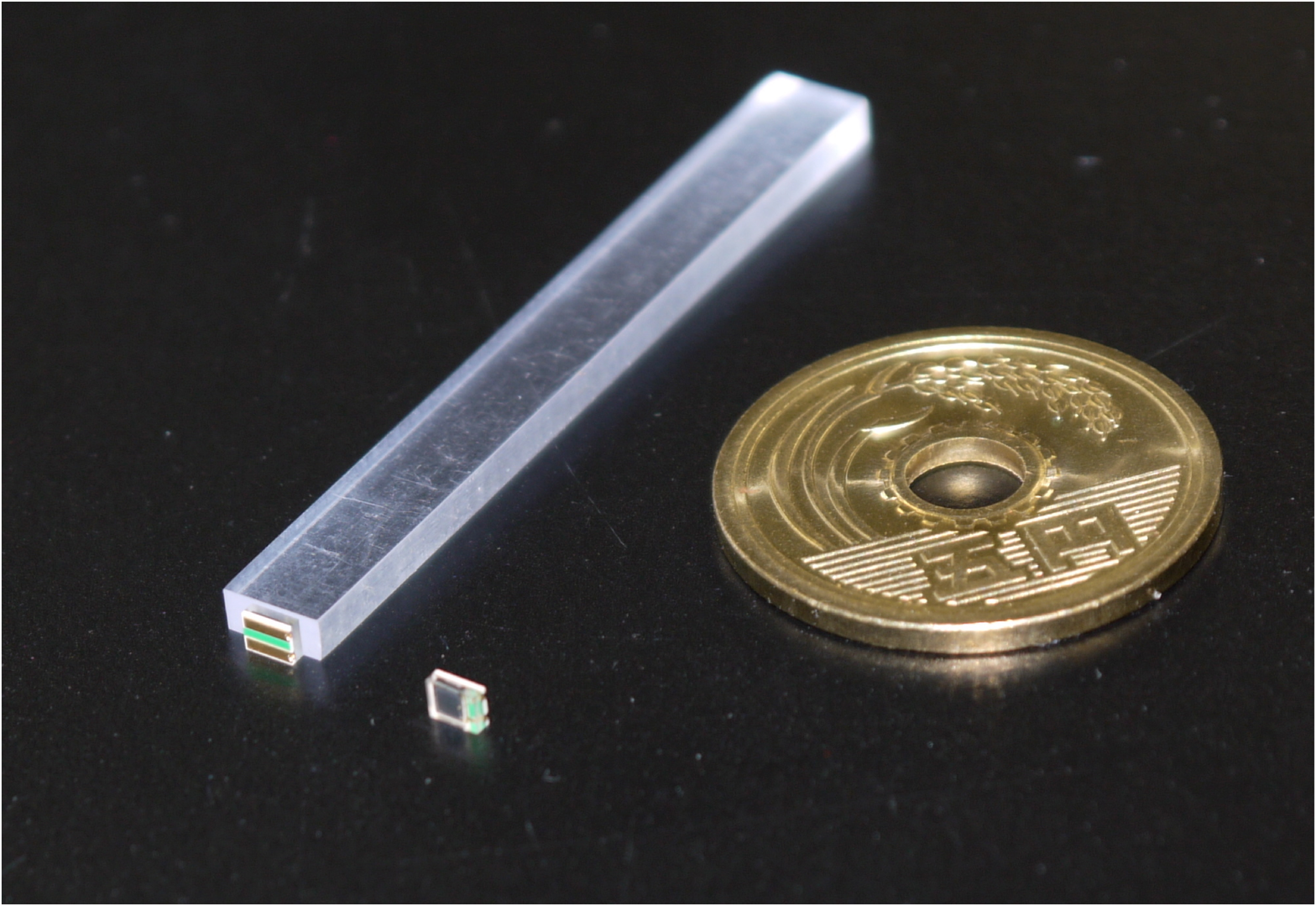}
\caption{\small {\it Left}: a view of ILD with a simulated multiple-jet event. 
The highlighted detector components are: 1. muon detector; 2. solenoid; 3. hadron calorimeter; 4. electromagnetic calorimeter; 5. TPC; and 6. vertex detector. 
{\it Right}: a 45\,mm\,$\times$\,5\,mm\,$\times$\,2mm scintillator strip and a PPD. 
} 
\label{fig:ILD} 
\end{center}
\end{figure}

Starting from the interaction point, the ILD consists of a vertex detector, silicon tracking layers, a large time projection chamber (TPC) surrounded by additional silicon tracking detectors,
a calorimeter system consisting of electromagnetic and hadronic sections, all placed within a solenoidal magnetic field of strength 3.5~T. The steel return yoke is instrumented
to provide muon identification. The basic structure consists of a central, ``barrel'', region aligned along the beam axis, closed by two ``endcaps'' in the forward regions.
The ECAL barrel detector has an octagonal cross-section, a length of around 5~m, and an inner radius of 1.85~m.
More details of the ILD design can be found in \cite{cite:DBD}. 
The ILD is simulated in \scmokka\,\cite{cite:mokka}, a \textsc{Geant4}-based simulation tool \cite{cite:geant}.
Figure \ref{fig:ILD} {\it left} shows a multiple-jet event simulated in ILD.

The ILD strip-scintillator ECAL (strip-ScECAL) is a sampling calorimeter. In the simulation model used in this study,
thirty sensitive layers are interleaved with tungsten plates of thickness 2.1 (4.2)~mm in the inner twenty (outer nine) layers.
The sensitive layers are tiled with scintillator strips of thickness 1~mm. Strips are aligned orthogonally in successive layers.
A dead volume of size $2.5 \times 0.91 \times 1.0$\,mm$^3$ is implemented at the edge of each scintillator to represent the volume occupied by the PPD.
In the case of strips shorter than 45\,mm, used as references, the PPD thick was scaled by the strip length to avoid large effect due to dead volumes.
Each strip is enveloped by a reflective film of thickness 57~$\mu$m.
Printed circuit boards housing the front end electronics and copper heat radiators are included in each layer.
The total thickness of such an ECAL is around 200~mm.

Four different scintillator tile configurations were used in this study:
\begin{enumerate}
	\item \fivefive\, tiles (``\mfivefive''); 
	\item \fortyfive\, strips (``\mfortyfive''); 
	\item alternating layers of \fivefive\, tiles and \fortyfive\, strips (``\altfive''); and
	\item alternating layers of \tenten\, tiles and \fortyfive\, strips (``\altten'').
\end{enumerate}
Successive strip layers were always orthogonally aligned.
An HCAL based on forty layers of 30\,mm\,$\times$\,30\,mm\,$\times$\,3\,mm scintillator tiles interleaved with 20~mm iron absorbers was simulated in this study.

\section{Strip Splitting Algorithm}\label{section:ssa}

\begin{figure}[h!] 
\begin{center}
\includegraphics[width=.9\textwidth]{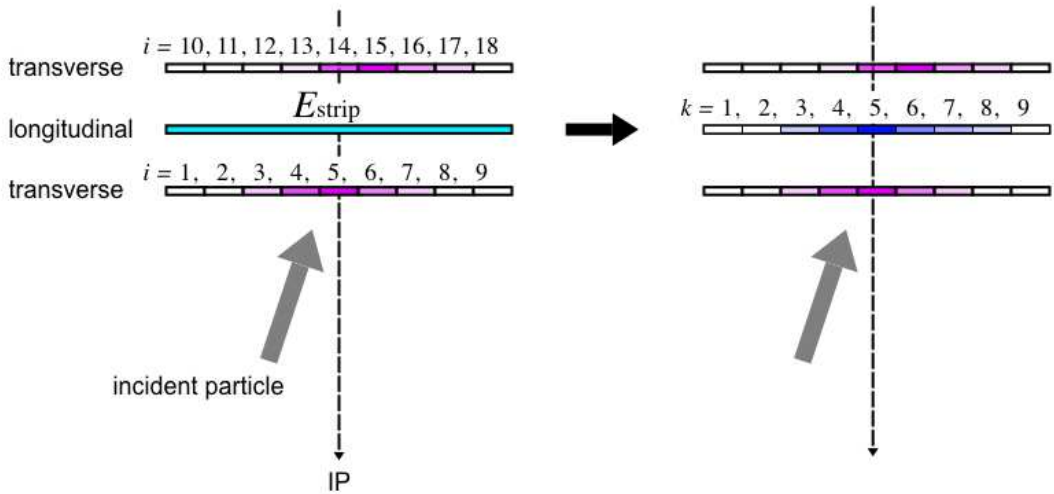}
\caption{\small
A cartoon illustrating the SSA procedure. 
The energy, $E_{\mathrm{strip}}$ reconstructed in the central, ``longitudinal'', strip is split among virtual cells ($k = 1, \cdots, 9$) by considering the energy in the 
orthogonally aligned ``transverse'' strips in neighboring layers.
More details of the procedure are given in the text.
 } 
\label{fig:SSAexplain} 
\end{center}
\end{figure}

A simple algorithm, the Strip Splitting Algorithm (SSA), has been developed to extract fine granularity information from long strip geometry.
Each strip is split into $n$ virtual cells along its length; $n$ is chosen to result in approximately square virtual cells, as an example, a 45\,$\times$\,5\,mm$^2$ strip is split into nine 5\,$\times$\,5\,mm$^2$ cells. 
The total energy $E_{\mathrm{strip}}$ detected by the strip is then distributed among the virtual cells  
according to the weights estimated by using the energy deposited on the strips in immediately neighboring layers, 
having intersection with the strip being considered, when seen from the interaction point of ILD.
Consequently, the energy deposited on the virtual cell $k$ is estimated as, 
\begin{equation}
	E_{\mathrm{virtual}}^k = E_{\mathrm{strip}}\cdot \Sigma_i E_{\mathrm{neighbor}}^{\{i,k\}}/ \Sigma_i E_{\mathrm{neighbor}}^i,
\end{equation}
where $k$ is the index of the virtual cell within the strip, $i$ is the index of neighboring intersecting strips, and $E_{\mathrm{neighbor}}^{\{i,k\}}$ is the energy deposited in the strip $i$ 
having intersection  in the range of virtual cell $k$.
Figure\,\ref{fig:SSAexplain} shows a schematic of the SSA procedure.

In the case of the \altten\ model, the \tenten\ tile is first split into $2\times2$ virtual cells. The neighboring strip layers are used to decide
how to partition the tile's energy among the virtual cells. In a second step, these virtual cells (originating from the \tenten\ tiles) are used to
decide the partitioning of energy among strips' virtual cells.
Although the use of \fivefive\ interleaving layers means that the presence of orthogonally aligned strip layers is no longer used, 
it is used again when tiles are larger than  \fivefive\ in this way.

Events were analyzed using a particle flow reconstruction algorithm. 
In the results presented later in this paper, the PandoraPFA algorithm\cite{cite:MarkT, cite:JohnCHEF} was used to analyze events,
together with other standard ILD reconstruction programs ({\it e.g.} for tracking) in the 
{\sc MarlinReco}\cite{cite:MARLIN} package.
PandoraPFA uses as input the energy and position of calorimeter deposits. When SSA was used, the centre of each virtual cell
and its assigned energy were used; when no SSA was used, the central position and total energy of each strip were used.

\section{Calibration}\label{section:calibration}

The calorimeter was calibrated by studying the energy deposited by 10~GeV photons for the electromagnetic response, and 10~GeV 
neutral long lived Kaons ($K_L$) for the hadronic response.
Particles were injected from the IP in a direction almost perpendicular to the beam-line (in order to avoid the central electrode of TPC)  
and uniformly distributed in azimuth. 
The calibration factors used to convert between the energy deposited in the scintillator and that deposited in the whole calorimeter were chosen to 
give a mean energy, after PFA reconstruction, equal to the incident particle energy.

Hadronic showers can start in ECAL, because the ScECAL has a nuclear interaction length of 0.8\,$\lambda$,
and electromagnetic showers can have tail in HCAL.
In those cases hadrons deposit the shower energy in ECAL with a ratio different from the electromagnetic shower, and vice versa.
Thus the correction factors $h/e$ are defined for ECAL and HCAL respectively and optimized to give a best jet energy resolution \cite{cite:f_h-eTune}. 

The ECAL was re-calibrated for each ECAL configuration, while a single HCAL calibration was used for all
configurations.


\section{Position of clusters}\label{section:position}

\begin{figure}[h!] 
\begin{center}
\includegraphics[width=.4\textwidth]{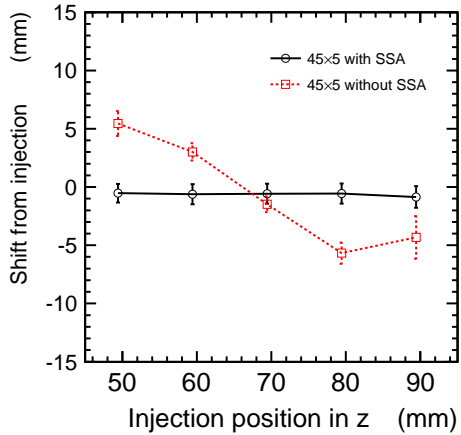}
\caption{\small Shift of reconstructed position for 10 GeV photons.} 
\label{fig:position} 
\end{center}
\end{figure}

The precise reconstruction of cluster positions is important in PFA analysis to ensure good matching between tracks and calorimeter clusters.
In this study, 10 GeV photons were fired into the 45$\times$5\, ECAL from the IP, 
varying the ECAL injection position along the $z$ direction, 
where we use a cylindrical coordinate system with its axis aligned along the beam line.
The injected position was taken to be the intersection on the ECAL front face of the line joining the reconstructed cluster position\footnote{
The position of a calorimetric cluster is 
defined as the energy-weighted mean position of its constituent hits.} and the IP. 
Figure\,\ref{fig:position} shows the difference between the reconstructed and true photon injection positions, as a function of the injected position.
The size of the vertical error bars reflects the width of the reconstructed position distribution.

When SSA is not used, a strong position-dependent bias of up to 5~mm is observed, corresponding to cases when the photon shower passes near the ends or centre of a strip
($z=69$~mm corresponds to the centre of strips aligned along the $z$ axis). When SSA is used, these biases are almost completely removed, and cluster positions are
reconstructed to better than 1~mm.

\section{Two-particle separation}\label{section:separation}

\subsection{$\mu$\,-\,$\mu$ separation}\label{section:mumu}

Di-muon events provide a sensitive system for measuring two-particle separation in different ScECAL designs.
Two like-sign muons of momentum 10 GeV 
with identical small differences in azimuthal and polar angles were injected into the detector at almost normal incidence to the beam-line.
The resulting simulated events were analyzed using SSA and PandoraPFA.

\begin{figure}[h!] 
\begin{center}
\includegraphics[width=.4\textwidth]{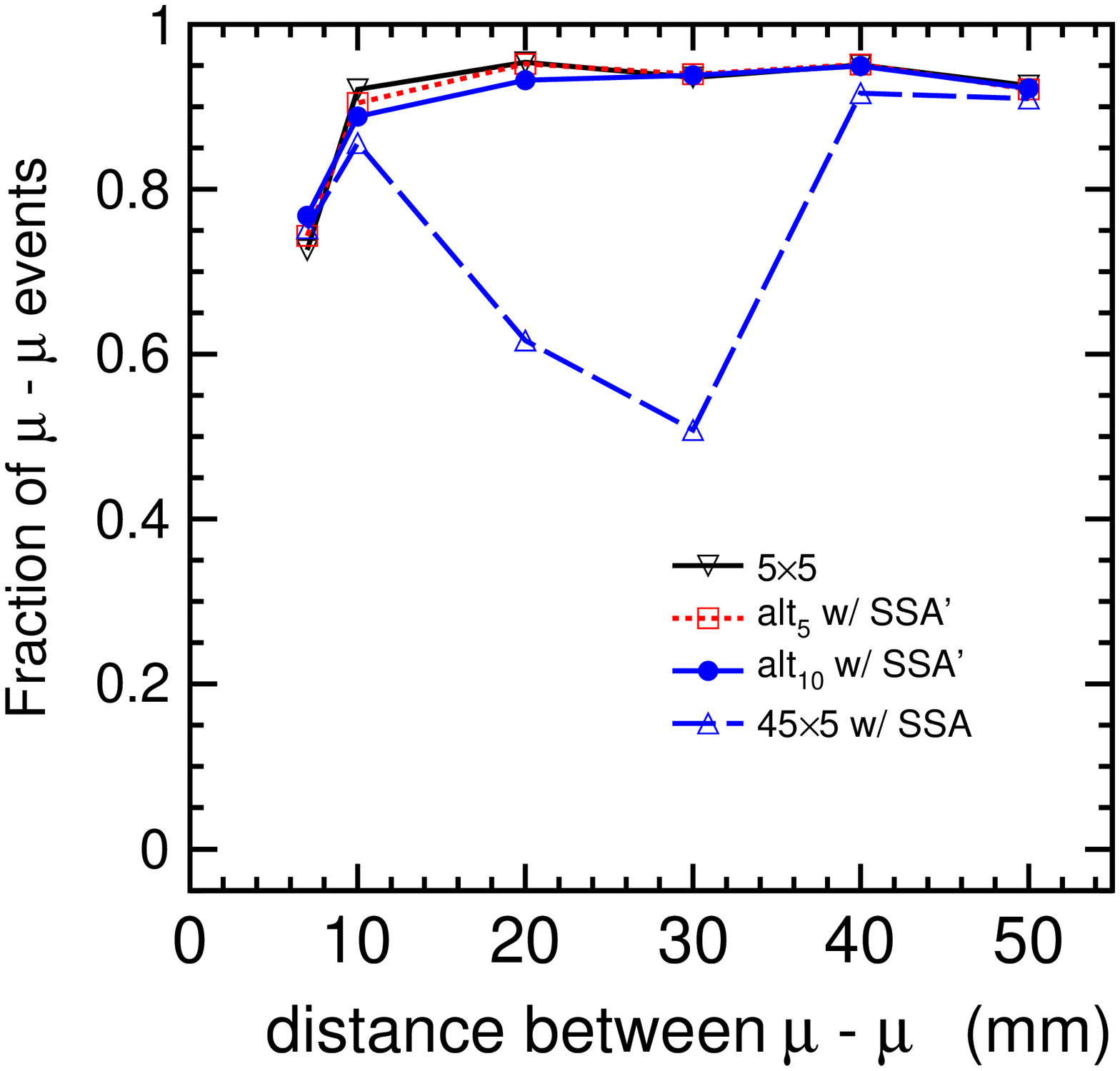}
\includegraphics[width=.35\textwidth]{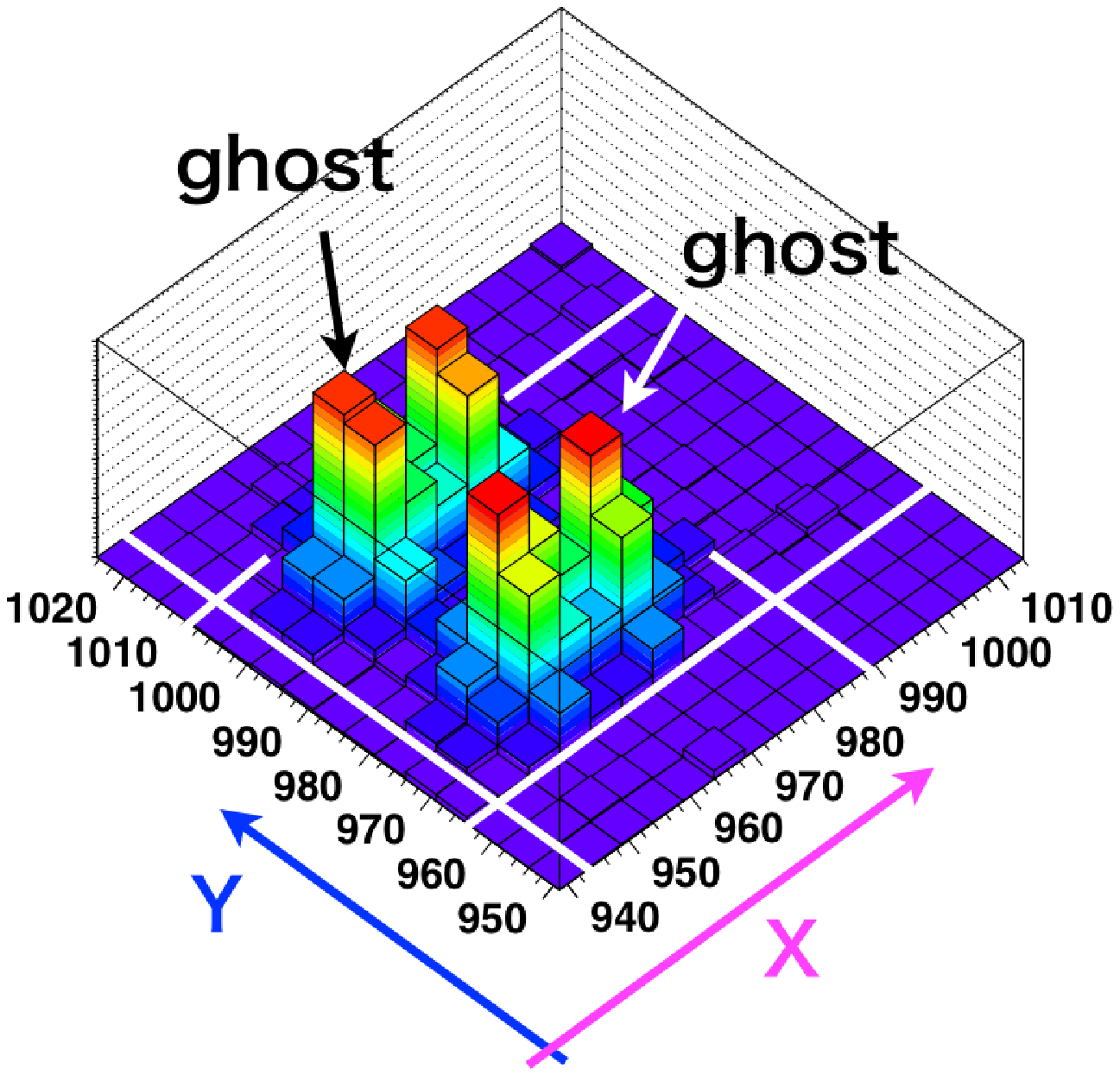}
\caption{\small {\it Left}: fraction of correctly reconstructed di-mion events as a function of the distance between the muons at the front face of the ECAL.   
{\it Right}: energy deposit of two electrons incident simultaneously on \mfortyfive\ strip-ScECAL reconstructed with SSA. Two of four peaks are ghosts.
} 
\label{fig:mumu} 
\end{center}
\end{figure}

Figure\,\ref{fig:mumu} {\it left} shows the fraction of such events in which two muons were correctly identified, as a function of the 
distance between the entry points of the two muons into the ScECAL. 
Apart from the \mfortyfive\,strip-ScECAL with the use of SSA, all types of ScECAL have
greater than 90\% of correctly reconstructed 
events when the distance between the two muons is larger than 10\,mm. 
In the case of the \mfortyfive\,strip-ScECAL with the use of SSA, the fraction depends 
strongly on the position, with a minimum at a distance of 30~mm. 
This is due to the formation of ghost tracks:
the possibility of two-fold ambiguities arising in reconstruction 
when two particles enter a square region with size smaller than the strip length.
Such cases can lead to the formation of ``ghost'' clusters.
Figure\,\ref{fig:mumu} {\it right} illustrates such a case in which two injected photons have produced two additional ghost clusters.
At a separation of 30\,mm, around 25\% of di-muon events in the  \mfortyfive\,strip-ScECAL have a single additional reconstructed ghost track,
and a further 25\% have two ghost tracks.
At distances below 30~mm, ghost tracks become more difficult
to resolve, and are more often combined with a true track.
At a separation of 7~mm, the correctly identified fraction drops to around 70\% for all ScECAL configuration due to 
the merging of nearby clusters.

The use of interleaved tile layers removes the ambiguities leading to ghosts, and dramatically improves the situation, 
giving a performance comparable to that of a tile-based ScECAL. 

\subsection{$\pi^0$ reconstruction to study two-photon separation}

A $\pi^0$ meson decays almost immediately into two photons, which can be reconstructed using only calorimeter information.
The $\pi^0$ energy has a strong influence on the opening angle between the two photons, and variations in the
decay angle give rise to different photon energies in the laboratory frame.
Sample of $\pi^0$ decays at different energies are therefore a powerful tool to measure ECAL performance, both in
terms of pattern recognition (the ability to identify two clusters), and energy resolution (by considering the
invariant mass of identified clusters).

\begin{figure}[h!] 
\begin{center}
\includegraphics[width=.4\textwidth]{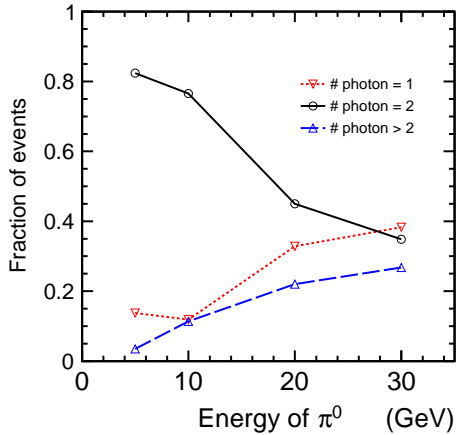}
\caption{
\small The fraction of the number of events which, when reconstructed in the \altten strip-ScECAL, contain one, two, and more than two reconstructed photons.
  } 
\label{fig:eventRatio} 
\end{center}
\end{figure}

Figure\,\ref{fig:eventRatio} shows the fraction of $\pi^0$ events in which
one, two, and more than two photons are reconstructed in the \altten\ ScECAL. 
The number of events with no reconstructed photons is less than 0.5\% for all $\pi^0$ energies. 
In around nine percent of events, at least one of the photons from $\pi^0$ decay converts before reaching the ECAL.
All ScECAL geometries, including the \mfivefive\,ScECAL, show similar behavior. This indicates that the increasing rate
of $\pi^0$ mis-reconstruction with energy is due to both the merging of photons into a single cluster and the fragmentation of photon clusters.

\begin{figure}[h!] 
\begin{center}
\includegraphics[width=.4\textwidth]{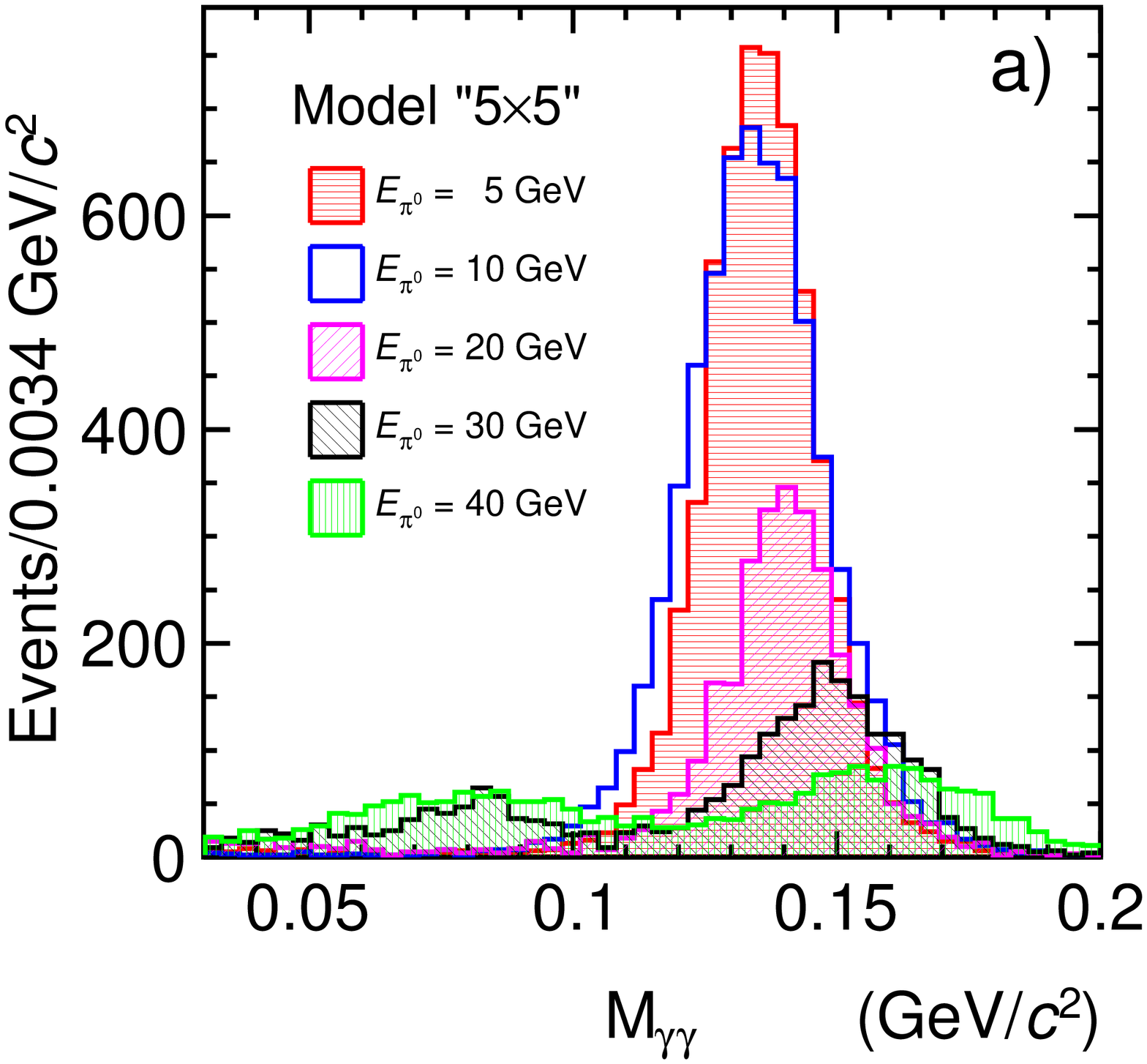}
\includegraphics[width=.4\textwidth]{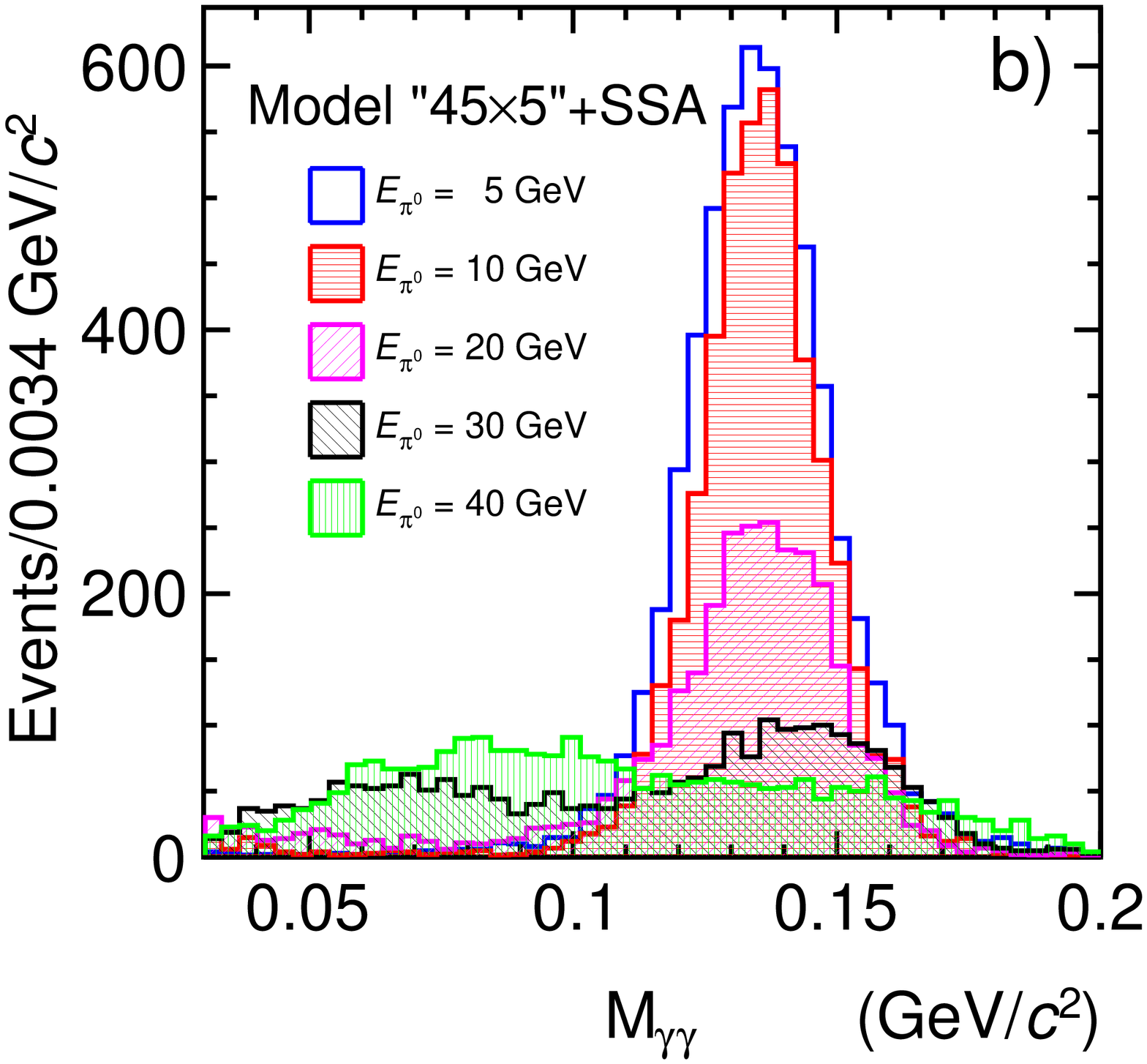}
\includegraphics[width=.4\textwidth]{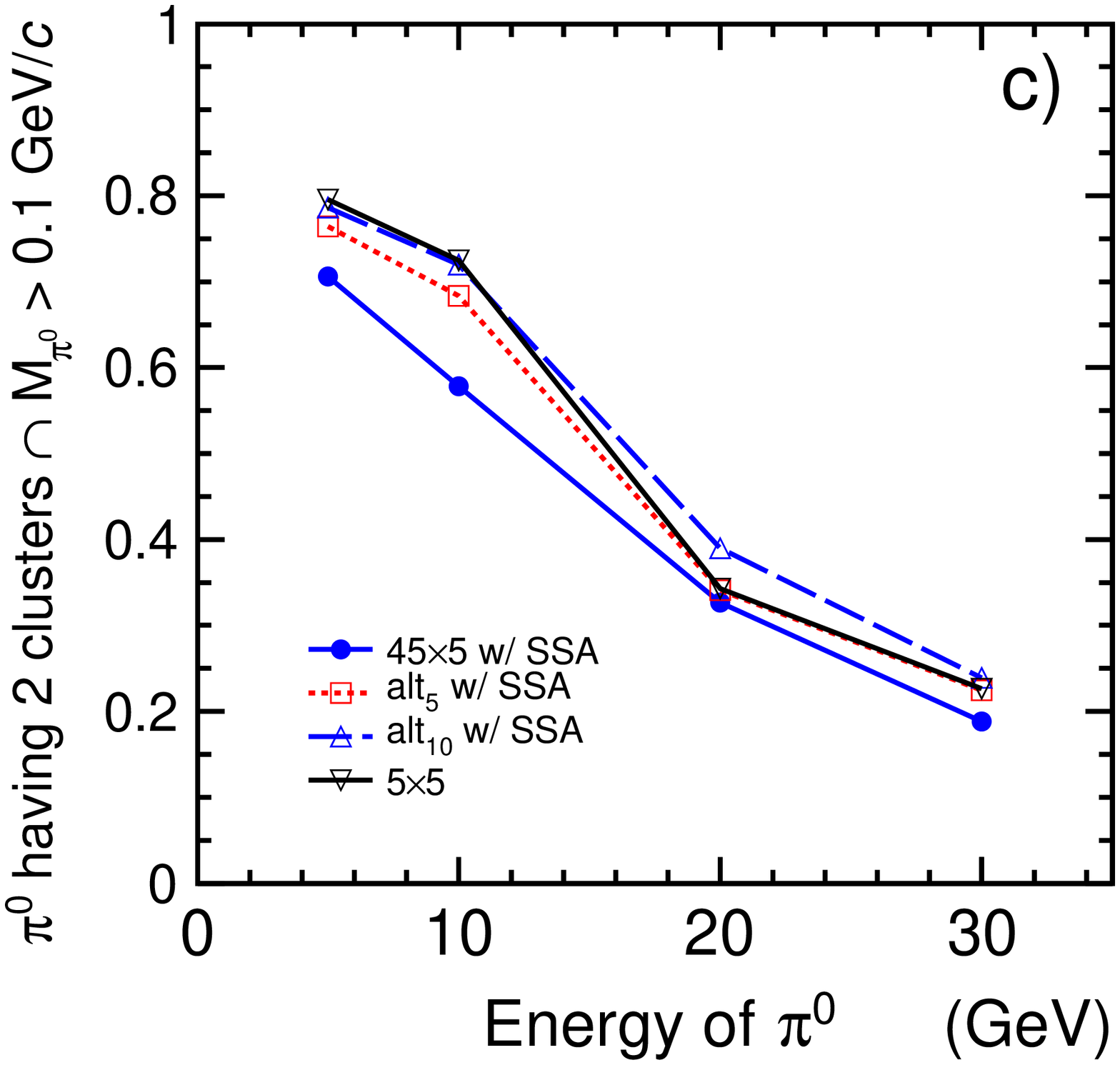}
\includegraphics[width=.4\textwidth]{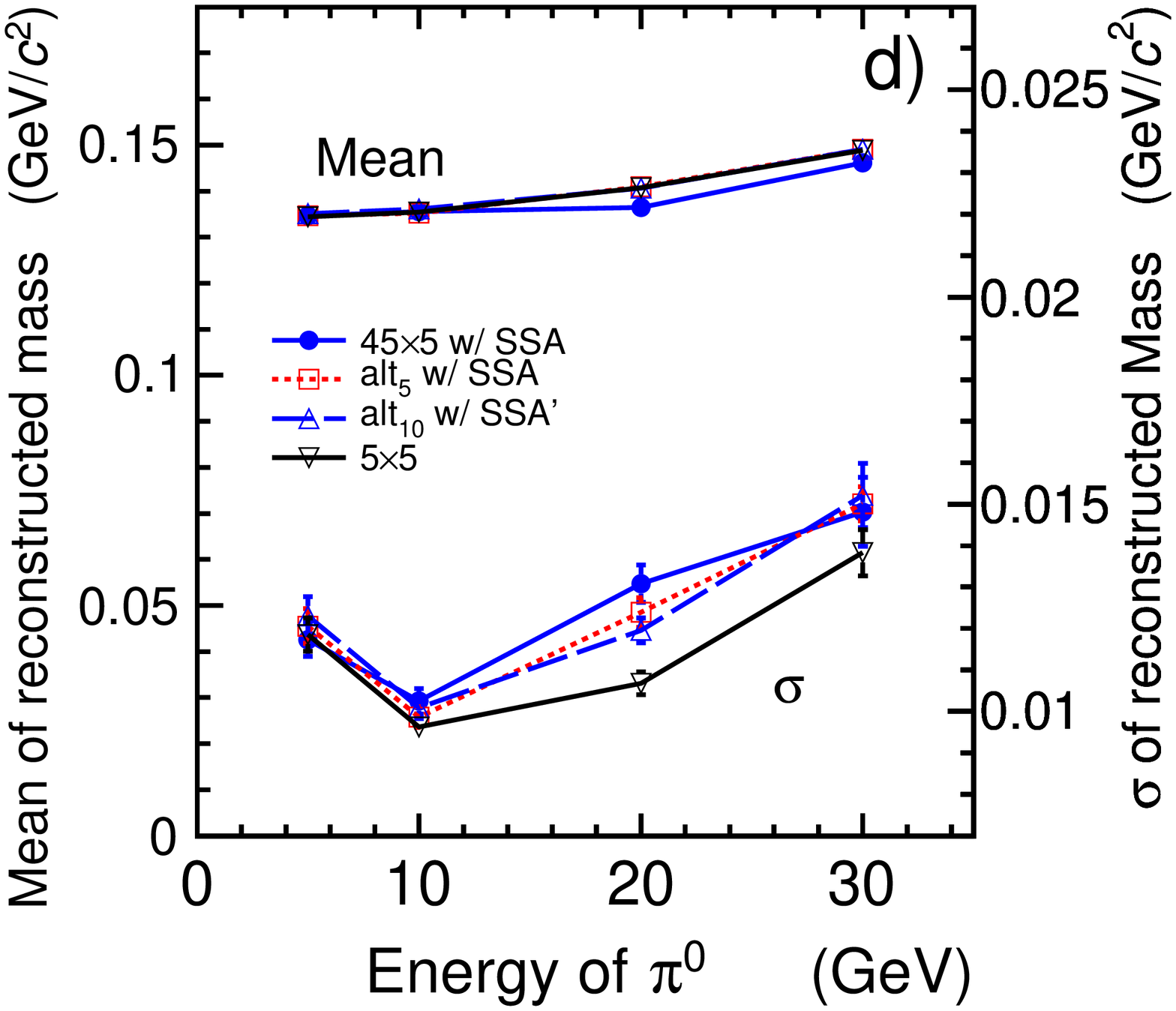}
\caption{\small Reconstructed \masspizero\ at different $\pi^0$ energies, for events 
with two reconstructed photon like clusters 
in (a) the \mfivefive\ tile-ScECAL, and (b) \mfortyfive\ strip-ScECAL,
and (c) the fractions of such events, in the four ECAL types, having M$_{\pi^0} > 0.1$\,GeV/$c^2$, and 
(d) the mean reconstructed \masspizero\ and resolution as a function of incident energy of $\pi^0$.
} 
\label{fig:massSpectrumPi0} 
\end{center}
\end{figure}

Figure \,\ref{fig:massSpectrumPi0} shows various characteristics of $\pi^0$ events in which exactly two photon-like 
clusters were identified. Figure\,\ref{fig:massSpectrumPi0}a (\ref{fig:massSpectrumPi0}b) shows the 
reconstructed invariant mass  of such events at different $\pi^0$ energies 
in a \mfivefive\ tile-ScECAL (\mfortyfive\ strip-ScECAL with SSA).
Figure\,\ref{fig:massSpectrumPi0}c shows the fraction of $\pi^0$ events which have 
exactly two identified photon-like clusters with a reconstructed invariant mass 
greater than 0.1 GeV/$c^2$, for the four ScECAL geometries considered.
The use of alternate tile layers improves the performance compared to a purely strip-based geometry, as was also
seen in the case of di-muon events.

Figure\,\ref{fig:massSpectrumPi0}d shows the means and standard deviations of Gaussian fits to 
the \masspizero\ spectra for $\pi^0$ of different energies, reconstructed in different ScECAL geometries.
SSA was used in all models except the \mfivefive\ tile-ScECAL.

These studies show that some level of $\pi^0$ reconstruction is possible at energies of up to $\sim$\,30\,GeV.
No large performance differences were seen between the four considered ECAL geometries.

\section{Jet energy resolution}\label{section:JER}
\subsection{Energy spectrum}

\begin{figure}[h!] 
\begin{center}
\includegraphics[width=.4\textwidth]{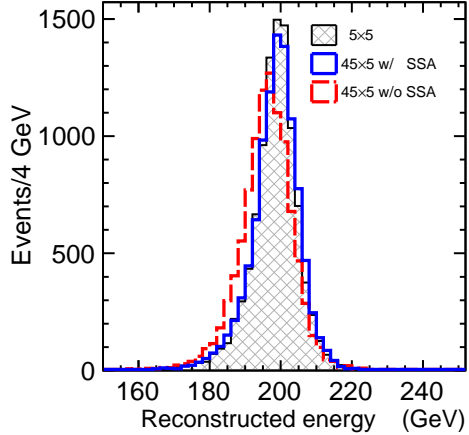}
\caption{\small Reconstructed energy in $q\bar{q}$ events at 200~GeV,  
for the \fivefive\ ScECAL and the  \fortyfive\ strip-ScECAL with and without the use of  SSA.
} 
\label{fig:250GeV} 
\end{center}
\end{figure}

The jet energy resolution was measured by analyzing $e^+ e^- \to q\bar{q}$ ($q=u,d,s$) events generated at centre-of-mass energies of 91.2, 200, 360, and 500~GeV.
These events were fully simulated in the ILD simulation described earlier, and reconstructed using {\textsc MarlinReco}, including SSA and PandoraPFA.
Figure\,\ref{fig:250GeV} shows the total reconstructed energy ($E_{jj}$) at a center-of-mass energy of 200 GeV when using the \mfivefive\ and \mfortyfive\ Sc-ECAL models,
with and without the use of SSA.
The shape of the energy spectrum for the \mfortyfive\ Sc-ECAL is noticeably improved by the use of SSA, and comes close to that of the \mfivefive\ model.

To estimate the jet energy resolution, the ``RMS90'' measure is used. 
Only events in the barrel region (absolute values of cosine of the angle between original quarks known from the simulation information and beam line $<$\,0.7) are considered, 
and ``RMS90'' is defined as the root mean square of the events
 lying within the smallest range of $E_{jj}$ which contains  90\% of events.

\subsection{Dependence on strip length}

\begin{figure}[h!] 
\begin{center}
\includegraphics[width=.4\textwidth]{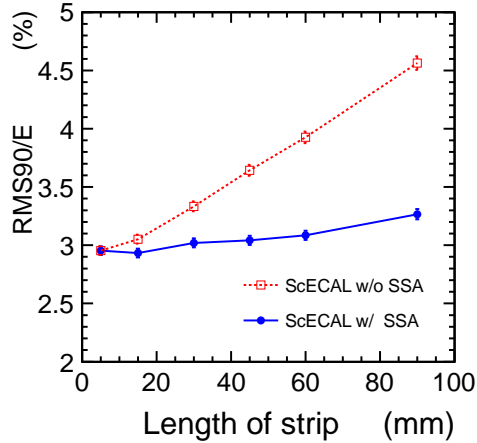}

\caption{\small Estimated jet energy resolution of 100\,GeV jets produced in $q\bar{q}$ events at a centre-of-mass energy of 200~GeV.} 
\label{fig:length} 
\end{center}
\end{figure}

The dependence of the jet energy resolution at a center-of-mass energy of 200\,GeV on the strip length is shown in Fig.\,\ref{fig:length}. 
The same strip width of 5~mm was used in all models.
The strong degradation in performance seen when SSA is not used is almost completely mitigated by the use of SSA.
The difference in jet energy resolution between an ScECAL using \mfivefive\ tiles and one using strips of length up to 60~mm is almost negligible.

\subsection{Comparison of \mfivefive, \mfortyfive, and \mfifteenfifteen\ ScECAL models}

\begin{figure}[h!] 
\begin{center}
\includegraphics[width=.4\textwidth]{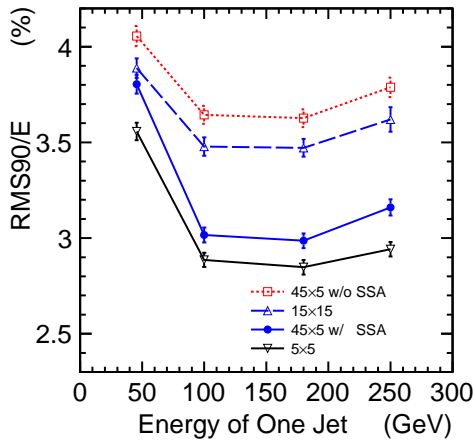}

\caption{\small Jet energy resolution of as  function of jet energy for different tile- and strip-ScECALs.
} 
\label{fig:15x15} 
\end{center}
\end{figure}

Figure\,\ref{fig:15x15} shows the jet energy resolution as a function of jet energy for a \mfortyfive\ strip-ScECAL without and with SSA, 
and \mfifteenfifteen\ and \mfivefive\ tile-ScECAL models.
At smaller jet energies, below 100 GeV, the jet energy resolution is dominated by the intrinsic single-particle resolution, 
giving a resolution which improves with increasing energy\footnote{
$\sigma_E/E$ of 3.5\% at 50\,GeV and 2.9\% at 100\,GeV is not consistent with $\sqrt{E}$ behavior.   
Probably, the confusion term starts to significant even below\,100 GeV.
}.
At energies above 100 GeV, the confusion between charged and neutral calorimeter clusters starts to become significant, 
leading to a degradation of resolution with increasing energy  \cite{cite:MarkT}.  
This behavior is seen for all ScECAL geometries.
The jet energy of the \mfortyfive\ strip-ScECAL is significantly improved by using SSA, especially at energies above 100 GeV, indicating that the confusion is decreased by SSA.
A tile of  \mfifteenfifteen\ has the same area as a \mfortyfive\ strip, however it is clear that the jet energy resolution when using the strip geometry
is significantly better (by up to 0.5\%), demonstrating the real merit of a strip-based geometry used in conjunction with SSA.
The degradation in jet energy resolution between the \mfivefive\ and \mfortyfive\ (with SSA) models is rather small, less than 0.25\% for 45~GeV jets, and
around 0.1\% for jets between 100 and 200~GeV. 

\subsection{Jet energy resolution of strip-tile-ScECAL}

As discussed in section \ref{section:mumu}, the major problem faced by a strip-based ECAL is  the formation of ghost clusters.
Therefore, the use of interleaving tile layers is expected to improve the jet energy resolution.

\begin{figure}[h!] 
\begin{center}
\includegraphics[width=.4\textwidth]{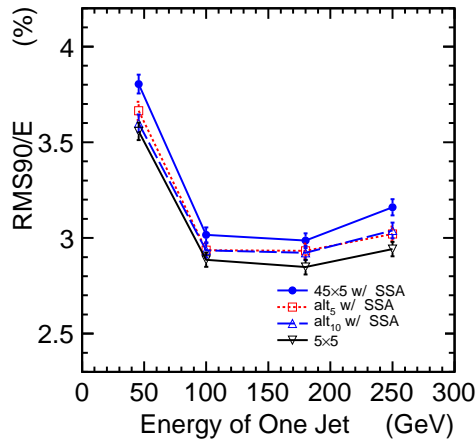}

\caption{\small Jet energy resolution as a function of jet energy for \mfortyfive, \altfive, \altten, and \mfivefive\ ScECAL models.  } 
\label{fig:strip-tile} 
\end{center}
\end{figure}

Figure\,\ref{fig:strip-tile} compares the jet energy resolutions of \mfortyfive, \altfive, and \altten\ models.
The performance of the \altfive\ and \altten\ models are almost identical. At jet energies of up to 100~GeV, they give almost the
same performance as the  \mfivefive\ model, while at higher energies the performance lies approximately half way between the
\mfivefive\ and \mfortyfive\ models.

\section{Discussion}\label{section:discussion}

SSA successfully extracts fine granularity information, at an effective scale close to that of the strip width, 
from a strip-based calorimeter.
The small degradation in jet energy resolution, around 0.2\%, when going from a  \mfivefive\ tile-ScECAL to a
\mfortyfive\ strip-ScECAL can be almost completely recovered by the use of tile layers interleaved between the strip layers.
These tile layers prevent the formation of ghost hits, as has been demonstrated in the reconstruction of a simple di-muon system.

Tile layers with a granularity of \tenten\ have been shown to work well. The use of such a tile size is 
technically feasible. The use of \mfifteenfifteen\,mm$^2$ tiles, which have the same area as the \fortyfive\ strips 
currently being used in a prototype ScECAL, and therefore the same density of readout electronics, 
are certainly technically feasible. Studies of reconstruction performance with such larger tiles are continuing.
The use of scintillator-based \fivefive\ layers is technically difficult at present, but a different technology, 
such as the silicon readout ECAL being developed by CALICE~\cite{cite:SiECAL}, could be used.

The strip scintillator technology for ECAL presented in this paper could also be applicable for the hadron calorimeter. 
It must be a challenging and worthwhile research to apply this method to more complicated topology of hadronic showers.

\section{Summary}\label{section:summary}

An algorithm (``SSA'') to extract fine granularity from scintillator strips was developed and tested.
The energy resolution for jets of up to 250\,GeV was compared among several types of ScECAL:
ScECAL with \fortyfive\,scintillator strips, and with such layers alternately replaced with \fivefive\, or \tenten\,tile layers.
Differences in the obtained jet energy resolutions when using a \mfivefive\,tile-ScECAL and \mfortyfive\,strip-ScECAL with SSA reconstruction 
ranged from 0.15\% to 0.25\%.
This difference can be removed for jet energies below 100\,GeV, or decreased to $\sim$\,0.1\% for jet energies in the range 150 - 250\,GeV, by alternately
 replacing strip layers with tile layers.
 
\section*{Acknowledgments}
The authors would like to thank to ILD group for providing essential simulation (\scmokka) and analysis tools (\scmarlin\  and PandoraPFA), and for useful discussions. John Marshall has given invaluable advice on the tuning and calibration of the PandoraPFA algorithm. Members of the CALICE collaboration, in particular the CALICE-ASIA group, have also contributed to many important discussions.
This work is supported in part by  `grant-in-aid for specially promoted research: a global research and development program of a state-of-the-art detector system for the ILC' of the 
Japan Society for Promotion of Science (JSPS).

 \end{document}